\documentclass[aps,prl,twocolumn,showpacs,preprintnumbers,superscriptaddress]{revtex4}
\usepackage{tabularx}
\usepackage{bm}
\usepackage{subfigure}
\usepackage{euscript}
\usepackage{epsfig,psfrag,subfigure}
\usepackage{graphicx,psfrag,subfigure}
\usepackage{color}
\usepackage{amsmath,amsfonts}
\usepackage{exscale}
\usepackage{amsbsy}
\usepackage{subfigure}

\def\be{\begin{equation}}
\def\ee{\end{equation}}
\def\ba{\begin{eqnarray}}
\def\ea{\end{eqnarray}}

\begin{document}

\title{Hydrodynamic description of transport
in strongly correlated electron systems}

\author{A. V. Andreev}
\affiliation{Department of Physics, University of Washington, Seattle, Washington
98195-1560, USA}

\author{Steven A. Kivelson}
\affiliation{Department of Physics, Stanford University, Stanford, California 94305, USA}

\author{B. Spivak}
\affiliation{Department of Physics, University of Washington, Seattle, Washington
98195-1560, USA}

\date{November 12, 2010}

\begin{abstract}
We develop a hydrodynamic description of the resistivity and magnetoresistance of an electron liquid in a smooth disorder potential. This approach is valid when the electron-electron scattering length is sufficiently short. In a broad range of temperatures, the  dissipation is dominated by heat fluxes in the electron fluid, and the resistivity is inversely proportional to the thermal conductivity, $\kappa$. This is in striking contrast with the Stokes flow, in which the resistance is independent of  $\kappa$ and proportional to the fluid viscosity. We also identify a new hydrodynamic mechanism of spin magnetoresistance.
\end{abstract}


\maketitle


Hydrodynamics accurately describes  most liquids at length scales long compared to the particle-particle mean-free path, $\ell$, but it is rarely relevant to the electron liquid in solids.  A hydrodynamic description is based on the existence of slow variables associated with conserved quantities, while neither the momentum nor the energy of the electron liquid in a solid is conserved;  electron-impurity ($ei$) and Umklapp scattering violate momentum conservation and electron-phonon ($ep$) scattering degrades both the momentum and energy of the electron fluid.  Consequently, even in relatively clean systems such as the electron gas in a semiconductor, the kinetics are typically described by the Boltzmann equation, and the conductivity is related to the corresponding momentum relaxation lengths, $\ell_{ei}$ and $\ell_{ep}$~\cite{Levinson}.

However, there are circumstances
in which the electron fluid in a semiconductor device, especially when it has very high mobility and is moderately strongly correlated, $r_s > 1$, exhibits a range of temperatures and sample purity where the electron-electron mean-free path, $\ell_{ee}$ is small compared to the length scales over which momentum conservation is violated, $\ell_{ee} \ll \ell_{ei}$ and $\ell_{ee} \ll \ell_{ep}$.  Moreover, under most circumstances, Umklapp scattering is negligible.  In this regime,  the electron fluid attains local equilibrium on the length scale $\ell_{ee}$, which is short compared to the scales at which the conservation laws break down, so the dynamics of the electron fluid can be treated hydrodynamically.  In this paper, we develop a theory of electron transport in the hydrodynamic regime.

Linear resistance of a solid object to a hydrodynamic flow was considered by Stokes long ago \cite{LandauLifshitz}. In this case the resistance is proportional to the first (shear) viscosity of the liquid $\eta$, and independent of the second viscosity $\zeta$ and the thermal conductivity $\kappa$, regardless of the compressibility of the liquid. The latter property can be traced to the fact that  the Stokes flow is \emph{isentropic}, i.e. the equilibrium entropy density of the fluid outside the obstacle is co-ordinate independent.  In the case of strongly correlated electronic systems this  is generally not the case. In the equilibrium state in the presence of a random potential the entropy per electron, $s_0({\bf r})$, is inhomogeneous, and consequently electron flow cannot be isentropic. We show below that in this case the resistance depends on all the kinetic coefficients: $\eta$, $\zeta$ and $\kappa$. Moreover, in the ideal fluid limit, $\kappa, \eta \to 0$, the resistivity diverges as $1/\kappa$, in contrast to the well known D'Alambert's paradox in Stokes flow. We also show that in the hydrodynamic regime the system exhibits a strong spin-dependent magneto-resistance.

While in the present paper we will not
analyze any explicit experimental system, there are reasons to believe that
our theoretical results may be relevant to existing experiments involving the highly correlated electron gas in semiconductor heterostructures.  Recently, low density two dimensional electronic systems with high mobility have become available, in which  $r_{s}=V/E_{F}\gg 1$ and the conductivity is relatively high even at low temperatures.  ($V$ is the characteristic energy of Coulomb interaction, and $E_{F}$ is the bare Fermi energy.)
These systems exhibit unusual temperature and magnetic field dependencies of the resistance. (See Refs.~\cite{kravchenko,KivelsonSpivak,KivelsonSpivak1} for a review.) If $r_{s}>1$ at $T\sim E_{F}$ the electron-electron mean free path is  as short as an  inter-electron distance so the hydrodynamic approach
should be applicable so long as the  correlation length of the scattering potential $\xi$, is large compared to the spacing between electrons.  (There are even indications that the hydrodynamic regime can be realized in GaAs MOSFET's at temperatures as high as $300 K$\cite{dyakonov,shurExp}.)

Having in mind the linear resistivity we use the Stokes approximation, which neglects the nonlinear terms in the hydrodynamic velocity. In the absence of external magnetic field, setting all time derivatives to zero (the stationary case), the Stokes equations for a charged fluid in the presence of an external potential are
\begin{subequations}\label{eq:Stokes}
\begin{eqnarray}\label{eq:continuity}
0 &=&  \boldsymbol{\nabla} \cdot \mathbf{j}  ,\\
  \label{eq:Navier_Stokes}
 0 & =& n_0^{-1}(\partial_k\sigma'_{ik}-\partial_i \tilde P) - (e E_i +\partial_i \tilde U\ ) ,\\
 \label{eq:entropy_flux}
0 &=& T \,  \mathbf{j}\cdot \boldsymbol{\nabla}  s_0  + \, \mathrm{div}\, \mathbf{Q}\ .
\end{eqnarray}
\end{subequations}
Here $T$, $n_0$ and $s_0$ denote respectively the temperature and \emph{position-dependent} particle  density and entropy per particle in equilibrium state;  $\mathbf{j}$ is  the particle current density; the electron charge is $-e$
and $\mathbf{E}$ is the homogeneous external electric field.
The linear in  $\mathbf{j}$ corrections to equilibrium quantities are indicated by the tilde sign: $\tilde P$ and $\tilde U$ are the current-induced  pressure and self-consistent potential. The latter is related to the nonequilibrium particle density, $\tilde n$, by the Poisson equation. The dissipative heat flux, $\mathbf{Q}$, and the viscous stress tensor, $\sigma'_{ik}$, are given by
\begin{subequations}\label{eq:constitutive}
\begin{eqnarray}\label{eq:viscous_stress}
\sigma'_{ik}&=& \eta \left( \partial_k v_i+\partial_i v_k -\frac{2}{d} \, \delta_{ik} \, \partial_l v_l\right ) + \zeta \partial_l v_l, \\
\label{eq:heat_flux}
\mathbf{Q}&=&- \kappa  \boldsymbol{\nabla} \tilde T.
 \end{eqnarray}
\end{subequations}
Here $\mathbf{v} = \mathbf{j}/n_0$ is the hydrodynamic velocity, $d$ the dimensionality of space, and $\tilde T$ is the nonequilibrium correction to the temperature due to the flow.
The hydrodynamic description of the resistivity  is fully determined by
Eqs.~(\ref{eq:Stokes}), (\ref{eq:constitutive}), augmented by the equation of state and the Poisson equation. The latter determine the spatial distribution of equilibrium density $n_0(\mathbf{r})$ and entropy $s_0(\mathbf{r})$, which are presumed known below.

To compute the resistivity $\rho$ one should solve
Eqs.~(\ref{eq:Stokes}-\ref{eq:constitutive}) and find the current response to the external electric field. Alternatively, for a given distribution of the current and temperature, the resistivity can be determined by equating the Joule heat, $\rho \, e^2 \,\langle \mathbf{j}\rangle ^2$,  to the dissipation rate of mechanical energy
\begin{equation}
\label{eq:entropy production_rate_currents_viscosity_neut}
    \rho \,e^2 \, \langle \mathbf{j} \rangle^2= \left \langle
    - \, \frac{  1}{T} \, \mathbf{Q} \cdot \boldsymbol{\nabla} \tilde T
    + \sigma_{ik}' \partial_k
    v_i
    \right\rangle,
\end{equation}
where $\langle \ldots \rangle \equiv \int \ldots d^d r /\int d^d r $ denotes averaging over space. This method is more convenient for perturbation theory and we will use it in this paper.

Let us begin with the simplest case of flow in one dimensional (1D) wire. The continuity equation (\ref{eq:continuity}) requires the current density to be uniform, $j=\langle j \rangle= const$, while the hydrodynamic velocity is given by $v=\langle j \rangle/n_0$. It follows from Eqs.~(\ref{eq:entropy_flux}) and (\ref{eq:heat_flux})  that the total heat flux
$q=Tj s_0-\kappa \partial_x \tilde T$  is uniform (independent of $x$).  Its value is established from the condition that the average temperature gradient vanishes, yielding $
\langle 1/\kappa \rangle\  q=[T \langle s_0/\kappa\rangle ] \ j$.
(In other words, the Peltier coefficient is $\Pi  =  -Q/e j =   - (T/e) \langle s_0/\kappa\rangle [\langle 1/\kappa \rangle]^{-1}$.) The temperature gradient is given by   $\partial_x \tilde T =(T/\kappa)  \delta s_0\ j$, where $\delta s_0 \equiv [s_0\langle 1/\kappa \rangle -\langle s_0/\kappa \rangle ]/ \langle 1/\kappa \rangle$. Substituting these expressions into Eq.~(\ref{eq:entropy production_rate_currents_viscosity_neut}) we obtain for the resistivity
\begin{equation}
\label{eq:resistivity_answer_1D}
  \rho_{1D}=  \frac{1}{ e^2}  \left\langle \frac{T }{\kappa} \left( \delta s_0\right)^2 +\zeta \left(
  \partial_x \frac{1}{n_0}\right)^2 \right\rangle.
\end{equation}
This equation assumes smooth variation of the disorder but does not
assume smallness of the relative variations of $s_0$ and $n_0$ and is valid in the case when the thermal conductivity and viscosity are also position-dependent.

The first term in Eq.~(\ref{eq:resistivity_answer_1D}) depends on the amplitude of spatial variations of $s_{0}(r)$ rather than their gradient. Therefore, for any given set of fluid parameters, the resistivity is dominated by thermal conductivity for a sufficiently smooth potential.

Another remarkable feature of Eq.~(\ref{eq:resistivity_answer_1D}) is that its first term is inversely-proportional to the thermal conductivity. This  implies that for an ideal fluid $\kappa \to 0$ the resistivity diverges, $\rho \to \infty$.  A qualitative explanation of the phenomenon is as follows. In equilibrium the entropy per electron  $s_{0}({\bf r})$ depends
only on the fluid density and becomes inhomogeneous in the presence of an external potential.  The flow of an ideal fluid is adiabatic and preserves the original entropy per particle. As a result the density dependence of the pressure will be different in different elements of the fluid. An adiabatic displacement of such a fluid from its equilibrium configuration will induce temperature gradients. The density change due to thermal expansion will create a restoring force, which is proportional to the displacement of the fluid, rather than to the gradient of the displacement. Consequently at $\kappa=0$ 1D adiabatic flow is impossible in linear response. This argument shows that the resistivity also diverges at $\kappa\to 0$ in two dimensions. It is easy to see from Eqs.~(\ref{eq:entropy_flux}) and (\ref{eq:heat_flux}) that adiabatic flow ($\kappa = 0$) is allowed only along contours of constant $s_0$, of which only a set of measure zero percolate across the system in 2D.

Equation (\ref{eq:resistivity_answer_1D}) is consistent with the result of a microscopic calculation of the resistivity of a 1D system of weakly interacting electrons, Eq.~(68) of Ref.~\onlinecite{Matveev2010}.

Let us now consider the resistivity in two dimensions (2D). In contrast to the 1D flow, in 2D current conservation does not uniquely determine the spatial dependence of the current density. It only implies that the latter can be expressed as
$\mathbf{j}=\langle \mathbf{j}\rangle +\hat z \times \boldsymbol{\nabla} \psi, $
where $\hat z$ is a unit vector perpendicular to the plane of flow, and the function $\psi$  describes spatial variations of the stream function~\cite{LandauLifshitz}.

The problem simplifies in the regime where the disorder potential is weak
so that perturbation theory can be applied.  In this case the relative fluctuations of equilibrium density and entropy are small and the viscosities and thermal conductivity may be assumed to independent of coordinates. Let us assume that the spatial fluctuations of the current density are much smaller than the average $ |\boldsymbol{\nabla} \psi | \ll |\langle \mathbf{ j} \rangle |$.
In this case the gradients of the hydrodynamic velocity and temperature are linear  in the inhomogeneity,
\begin{equation}\label{eq:gradients_PT}
    \partial_i v_k = \langle j_k \rangle \partial_i (n_0^{-1}), \quad \mathbf{\nabla} \tilde T =\frac{T}{\kappa} \langle \mathbf{j} \rangle  \delta s_0.
\end{equation}

Substituting these expressions into Eq.~(\ref{eq:entropy production_rate_currents_viscosity_neut})  we obtain for the resistivity in the 2D case,
\begin{equation}
\label{eq:resistivity_answer_2D}
  \rho_{2D}=  \frac{1}{ 2 e^2}  \left\langle \frac{T }{\kappa} \left( \delta s_0\right)^2 +(\eta +\zeta) \left( \nabla \frac{1}{n_0}\right)^2 \right\rangle.
\end{equation}
Although this expression looks similar to the 1D result, Eq.~(\ref{eq:resistivity_answer_1D}), this formula is applicable only to weakly inhomogeneous flows. Therefore all kinetic coefficients are coordinate-independent and correspond to those of the disorder-free state, and $\delta s_0 =s_0-\langle s_0 \rangle$.
The origin of the first two terms in
Eq.~(\ref{eq:resistivity_answer_2D}) is the same as that in Eq.~(\ref{eq:resistivity_answer_1D}).  The additional factor of $1/2$ corresponds to the inverse number of dimensions. It arises because only the gradients of $s_0$ and $n_0$ along the flow contribute to the resistance. The term containing the shear viscosity $\eta$ in Eq.~(\ref{eq:resistivity_answer_2D}) arises because in 2D the inhomogeneous part of the hydrodynamic velocity contains shear flow.

It is interesting to compare this expression with the classical
expressions for the resistance to a flow past a set of fixed objects.  In this case, rather than the smoothly varying disorder we have treated, one considers the hydrodynamic flow in the presence of spatially sharp objects, along the surface of which ``stick'' boundary conditions (${\bf v}({\bf r})={\bf 0}$) are applied on the hydrodynamic velocity. Under these conditions, the Stokes formula for the resistance produced by a set of macroscopic objects embedded into a 2D liquid is
\begin{equation}\label{stocks}
\rho \sim \frac{1}{e^2} \frac{\eta N_i}{n^2 |\ln R^2 N_i|}.
\end{equation}
Here $N_{i}$ is the concentration of objects and $R$ is their radius.
Clearly, this expression is similar to that obtained from the second term in Eq.~(\ref{eq:resistivity_answer_2D}), although our expression contains the sum $(\eta+\zeta)$, while Eq.~(\ref{stocks}) contains only $\eta$.
However, more importantly, as in the 1D case, when the disorder is sufficiently 
 {smooth}, the resistance is dominated by the thermal conductivity contribution, described by the first term in Eq.~(\ref{eq:resistivity_answer_2D}).

The perturbative result, Eq.~(\ref{eq:resistivity_answer_2D}), applies
when the flow is nearly homogeneous. If the thermal conductivity $\kappa$ decreases a strongly inhomogeneous flow will develop even if the equilibrium density is almost uniform, $| \delta n_0/n_0| \ll 1$. This is obvious in the $\kappa \to 0$ limit, where the linear response flow is possible only along the $s_0=const$ lines. At finite but small thermal conductivity the current flows primarily in narrow channels  localized near the contours of constant $s_0$ that percolate across the whole sample.  We can estimate the width of the channels $x$ and hence $\rho_{2D}$ in this limit by computing  the rate of energy dissipation in Eq.~(\ref{eq:entropy production_rate_currents_viscosity_neut}) for an assumed value of $x$ and then minimizing with respect to $x$.
The first term in Eq.~(\ref{eq:entropy production_rate_currents_viscosity_neut}) may be estimated as   $\frac{\eta \langle j \rangle^2}{\langle n_0 \rangle^2 \xi^2} (\xi/x)^3 $ and the second as $\frac{T \langle (\delta s_0)^2 \rangle }{\kappa} \langle j \rangle^2 (x/\xi)^2$, where $\xi$ denotes the correlation length of the disorder potential. Minimizing the sum gives
$x\sim \xi \alpha^{-1/6}$,
where
\begin{equation}
\alpha \equiv \frac{\langle \delta s_0^2\rangle  T \xi^2 \langle n_0 \rangle^2}{\kappa \eta }  \gg 1  .
\end{equation}
The corresponding resistivity of the sample is
\begin{equation}\label{eq:resistivity_inhomogeneous}
 \rho_{2D}\sim \frac{1}{e^2}  \sqrt{\frac{T \eta \langle (\delta s_0)^2 \rangle}{\kappa \langle n_0 \rangle^2 \xi^2}}=\frac {\eta}{e^2\langle n_0\rangle^2\xi^2} \sqrt{\alpha} .
\end{equation}
Note that this limit is likely relevant at relatively high temperatures,
when $\kappa$ is small, although not so high that the inequality
$\tau_{ee}\gg \tau_{ep}$ is violated. In the opposite limit $\alpha \ll 1$ the flow is nearly homogeneous, and the resistivity is described by the perturbative result, Eq.~(\ref{eq:resistivity_answer_2D}).

We now consider the generalization needed to explore a spin mechanism of magnetoresistance. We assume that the component of spin parallel to an applied field  $\mathbf{H}$ is conserved, so the spin \emph{per electron}, $\sigma$, is a new hydrodynamic variable.  For simplicity, we ignore the orbital effects of the magnetic field.

The hydrodynamic equations in the present case should be supplemented by
the conservation law for spin, which in the
linear in $\mathbf{j}$ approximation reads
\begin{equation}\label{eq:spin_current_conservation}
\mathbf{j}\cdot \boldsymbol{\nabla} \sigma_0=  -\, \mathrm{div} \, \mathbf{j}_\sigma.
\end{equation}
Here $\sigma_0(\mathbf{r}, H)$ is the equilibrium value of $\sigma$ and
$\mathbf{j}_\sigma$ is the density of spin current relative to the fluid. The latter consists of a spin diffusion current and a spin thermo-current induced by the temperature gradients.  Conversely, the heat flux $\mathbf{Q}$ acquires an additional contribution in the presence of nonequilibrium spin density gradients, as required by the Onsager principle. We can express the heat and spin currents in terms of the kinetic coefficients and gradients of temperature and ``spin chemical potential'' $\mu_\sigma$ as
\begin{equation}\label{eq:heat_spin_flux}
    \left[
      \begin{array}{c}
        \mathbf{Q} \\
        \mathbf{j}_\sigma \\
      \end{array}
    \right] =   -\,\frac{1}{T}\left(
                \begin{array}{cc}
                  \gamma_{11} & \gamma_{12} \\
                  \gamma_{21} & \gamma_{22} \\
                \end{array}
              \right) \left[
                        \begin{array}{c}
                         \boldsymbol{\nabla} \tilde T/T \\
                         \boldsymbol{\nabla} \mu_\sigma  \\
                        \end{array}
                      \right].
\end{equation}
Here and
$\gamma_{12}(H)=\gamma_{21}(H)$, and the diagonal elements can be expressed in terms of the thermal conductivity $\kappa$ and spin diffusion coefficient $D_\sigma$ as $\gamma_{11}=T^2 \kappa$, and  $\gamma_{22}=D_\sigma$

The expression for the resistivity in terms of the dissipation rate,  Eq.~(\ref{eq:entropy production_rate_currents_viscosity_neut}), must now be replaced with
\begin{equation}\label{eq:entropy production_rate_spin}
    \rho e^2 \langle \mathbf{j} \rangle ^2= \left\langle \sigma_{ik}'\frac{\partial
    v_i}{\partial x_k}- \frac{\mathbf{Q} \cdot
    \boldsymbol{\nabla } \tilde T}{T} - \,
    \mathbf{j}_\sigma \cdot \boldsymbol{\nabla }\tilde \mu_\sigma
    \right\rangle.
\end{equation}

In the weakly inhomogeneous regime we have (in vector notation) $(\mathbf{Q}, \mathbf{j}_\sigma ) =\, -\, \mathbf{j} \, (T \delta s_0, \delta \sigma_0) $, where $\delta \sigma_0 =\sigma_0 -\langle \sigma_0 \rangle$. Using Eqs.~(\ref{eq:heat_spin_flux}) and (\ref{eq:entropy production_rate_spin}) we obtain the resistivity
\begin{equation}
\label{eq:resistivity_answer_spin}
 \rho_{2D} =
   \frac{T}{ 2 e^2} \! \left\langle\!
\left(\!
  \begin{array}{cc}
    T \,\delta s_0,\!& \!\delta \sigma_0 \\
  \end{array}
  \!\right)
  \hat \gamma^{-1} \!\left(\!
                                 \begin{array}{c}
                                   T \, \delta s_0 \\
                                   \delta \sigma_0 \\
                                 \end{array}
                               \!\right)\!
                               + \!\frac{\eta +\zeta}{T} \!\left[ \nabla \frac{1}{n_0}\right]^2\!
                               \right\rangle,
\end{equation}
where $\hat{\gamma}^{-1}$ is the  matrix inverse to $\hat{\gamma}$.
The magetoresistance arises not only from the $ H$-dependence of the kinetic coefficients $\eta(H), \zeta(H)$ and $\hat{\gamma}(H)$, but also
from the equilibrium quantities $\delta s_{0}( H)$ and $\delta \sigma_{0}( H)$. Spin polarization by the magnetic field decreases $\delta s_{0}$, thereby decreasing thermal dissipation. On the other hand, it induces spatial inhomogeneity of the convective spin current $\mathbf{j}\sigma_{0}({\bf r})$. This generates diffusive spin currents, which increase the resistance of the sample.

The hydrodynamic results are very general -- the physics of the particular electron fluid involved enters only through the magnitude and functional dependences of the various kinetic coefficients on the magnetic field and temperature. In the case  of $r_{s} \lesssim 1$,  theoretical calculations of
these coefficients  are under good control. (See, for example, Ref.~\cite{LifshitzPitaevskii}.) For $T<E_{F}$, where Fermi liquid theory applies, $\delta s_0
\sim (T/E_{F}) \delta n_0 /n_0$, $\eta\sim \zeta\sim m E_F^3/\hbar T^2 \ell_{ee}$, and $\kappa\sim E_F^2/\hbar T$,  where $m$ is the electron mass. The first term in Eq.~(\ref{eq:resistivity_answer_2D}) is on the order $\frac{\hbar}{e^2} \frac{T^4}{E_F^4}\langle \delta n_0^2/n_0^2 \rangle$ and the second term $\sim \frac{\hbar}{e^2} \frac{1}{\xi^2 n_0} \frac{E_F^2}{T^2} \langle \delta n_0^2/n_0^2 \rangle$, where $\xi$ is the correlation length of the disorder potential. At temperatures close to the Fermi energy  the thermal conductivity contribution in Eq.~(\ref{eq:resistivity_answer_1D}) is larger than the viscous one by a factor $\xi^2 n_0 \gg 1$,
so $\rho \sim \frac{\hbar}{e^2} \langle \delta n_0^2/n_0^2 \rangle$.
For $T > E_F$, the electrons form a classical gas, so
$\delta s_0 \sim \delta n_0/n_0$, $\kappa \sim v_T T/e^2$, $\eta \sim \zeta \sim m v_T T/e^2$, where $v_T \sim \sqrt{T/m}$ is the thermal velocity.
The thermal conductivity term in Eq.~(\ref{eq:resistivity_answer_2D}) decreases with temperature $\sim \frac{1}{v_T}\langle \delta n_0^2/n_0^2 \rangle$ and the viscous one increases, $\sim \frac{1}{v_T}\langle \delta n_0^2/n_0^2 \rangle \frac{1}{\kappa_D^2 \xi^2}$, where $\kappa_D\sim n_0 e^2/T$ is the inverse Debye screening length.

In clean  systems at $T=0$ the electron fluid crystallizes for large enough values of $r_{s} > r_{s}^{(c)}\gg 1$. In 2D case, the best estimate of $r_{s}^{(c)}\sim 40$ has been obtained by numerical simulations under the assumption that there is a direct transition between the crystal and the liquid states  \cite{Ciperley}. Although it has been shown \cite{KivelsonSpivak,KivelsonSpivak1} that, rather than a direct transition, there must be a sequence of transitions involving electronic microemulsion phases, it still seems likely  that there is a broad interval $r_s^{(c)}> r_{s}\gg 1$ where at $T=0$  the system is in the liquid state. In this case, there is a more involved hierarchy of crossover scales, since $E_{F}^{\star}\ll \Omega_{P}\ll V$, where $E_F^{\star}$ is the (probably strongly renormalized, $E_F^\star < E_F$) Fermi energy, $\Omega_{p}\sim E_F \sqrt{r_s}$ is the plasma frequency at wave vector or order $n^{1/2}$, and $V\sim E_F r_s$ is the typical interaction energy between electrons. As a result, there are four characteristic temperature intervals:
1)  For $T < E_{F}^\star$ the system is in the Fermi liquid regime, which behaves as above.
2)  For $E_{F}^\star < T < \Omega_{p}$ there is a semi-quantum regime.
While there is no established theory in this regime, a conjecture concerning the $T$- dependences of $s(T)$, $\kappa$, $\eta$, and $\zeta$
was put forward in Refs.~\cite{andreev,andreev1,KivelsonSpivak}.
Surprisingly, there also does not appear to be published experimental data
concerning the transport properties of liquid $^{3}He$ in the corresponding regime, much less for the strongly correlated electron liquid.  3)  For $\Omega_{p}<T<V$ the system is a highly correlated classical fluid, a classic problem about which much is known empirically, but which is still a subject of ongoing theoretical debate~\cite{gilles}.  4)  For $V < T$, the system is again weakly interacting classical plasma.

\acknowledgements

The authors would like to thank K. A. Matveev and Yu. Nazarov for useful discussions. This work was supported by the DOE Grant No. DE-FG02-07ER46452 (AVA), and the NSF Grant Nos. DMR-0704151 (BS) and DMR-0758356 (SAK).

\end{document}